# Unveiling hidden structures in the Coma cluster *

A. Biviano[1], F. Durret[2,3], D. Gerbal[2], O. Le Fèvre[3], C. Lobo[2,4], A. Mazure[5], and E. Slezak[6]

[1] Sterrewacht Leiden, Postbus 9513, Niels Bohrweg 2, 2300 RA Leiden, The Netherlands
[2] Institut d'Astrophysique de Paris, CNRS, Université Pierre et Marie Curie, 98bis Bd Arago, F-75014 Paris, France
[3] DAEC, Observatoire de Paris, Université Denis Diderot, CNRS (UA 173), F-92195 Meudon Cedex, France
[4] Centro de Astrofísica da Universidade do Porto, Rua do Campo Alegre 823, 4100 Porto, Portugal
[5] Laboratoire d'Astrophysique Spatiale, CNRS, Traverse du Siphon, Les Trois Lucs, B.P. 8, F-13376 Marseille Cedex, France
[6] Observatoire de la Côte d'Azur, B.P. 229, F-06304 Nice Cedex 4, France

**Abstract.** We have assembled a large data-set of 613 galaxy redshifts in the Coma cluster, the largest presently available for a cluster of galaxies. We have defined a sample of cluster members complete to $b_{26.5} = 20.0$, using a membership criterion based on the galaxy velocity, when available, or on the galaxy magnitude and colour, otherwise. Such a data set allows us to define nearly complete samples within a region of 1 $h^{-1}$ Mpc radius, with a sufficient number of galaxies per sample to make statistical analyses possible. Using this sample and the *ROSAT* PSPC X-ray image of the cluster, we have re-analyzed the structure and kinematics of Coma, by applying the wavelet and adaptive kernel techniques.

A striking coincidence of features is found in the distributions of galaxies and hot intracluster gas. The two central dominant galaxies, NGC 4874 and NGC 4889, are surrounded by two galaxy groups, mostly populated with galaxies brighter than $b_{26.5} = 17$ and well separated in velocity space. On the contrary, the fainter galaxies tend to form a single smooth structure with a central peak coinciding in position with a secondary peak detected in X-rays, and located between the two dominant galaxies; we suggest to identify this structure with the main body of the Coma cluster. A continuous velocity gradient is found in the central distribution of these faint galaxies, a probable signature of tidal interactions rather than rotation. There is evidence for a bound population of bright galaxies around other brightest cluster members.

Altogether, the Coma cluster structure seems to be better traced by the faint galaxy population, the bright galaxies being located in subclusters. We discuss this evidence in terms of an ongoing accretion of groups onto the cluster.

## 1. Introduction

The Coma cluster of galaxies, number 1656 in the Abell (1958) catalogue, has been one of the most studied galaxy clusters since the first determination of its mass by Zwicky (1933). Its high richness (richness class 2) and smooth appearance made of this cluster the prototype of well relaxed clusters, in contrast to the most nearby cluster, Virgo, considered as the prototype of irregular clusters. Support to this view came from the discovery by Rood (1969) of luminosity segregation among the galaxies in Coma, interpreted as an indication of partial dynamical relaxation.

However, as early as in the 70's some facts started to challenge the picture of Coma as a well relaxed dynamical system: Rood & Sastry (1971) classified Coma as a "binary" cluster, because of the presence of a couple of giant galaxies in its core (NGC 4889 and NGC 4874), and Rood (1974) hinted to a possible subclustering around these two galaxies. Valtonen & Byrd (1979) went even farther, by proposing an unconventional dynamical model in which most of the cluster mass was linked to these two giant galaxies. During the 80's the claims for subclustering in Coma became more and more frequent: Baier (1984) claimed significant substructure in the South-West region of Coma, around the giant galaxy NGC 4839, and hinted at a possible subclustering in the cluster core; Gerbal et al. (1986) and Mellier et al. (1988) suggested that subclustering occurs around many bright galaxies and is not only limited to NGC 4889, 4874 and 4839; Fitchett & Webster (1987) used the Lee-statistics to provide the first significant detection of subclustering in the cluster core.

Nevertheless, in the same period, Dressler & Shectman (1988) did not find a significant signal for subclustering in Coma, and most dynamical analyses still assumed a smooth cluster structure (Kent & Gunn 1982; The & White 1986; Merritt 1987; Hughes 1989). It was in fact not until the 90's that the complicated structure of the Coma cluster became widely accepted, thanks to new techniques applied to the galaxy distribution (Escalera et al. 1992) and to the detection of these structures in the X–ray band with the *Spacelab 2* X-ray telescope (Watt et al. 1992) and particularly with *ROSAT* (Briel et al. 1992; White et al. 1993; Vikhlinin et al. 1994).

Subclustering implies an incomplete virialization of the cluster, with ongoing mergers of groups onto the main body of the cluster; the dynamical youth of the Coma cluster is also indicated by the kinematical difference between the star-forming galaxies and the old galaxy population (Sodré et al. 1989; Zabludoff & Franx 1993; Gavazzi et al. 1995; Colless & Dunn 1995, hereafter CD95); the existence of a population of E+A spectrum galaxies likely to have suffered a burst of star formation within the last 2 Gyr (Caldwell et al. 1993, 1995) is also suggestive of ongoing accretion of galaxies onto the cluster. Finally, unlike most clusters, Coma lacks a cooling flow (Edge et al. 1992) and has a radio halo (Hanisch 1980); these features have been interpreted in terms of an ongoing collision of two equally sized subclusters (McGlynn & Fabian 1984; Edge et al. 1992; Tribble 1993; Fabian et al. 1994).

Coma can now be considered as the prototype of rich clusters endowed with subclusters, and thus not fully relaxed. However, the very presence of substructures logically implies the existence of a *main body*, on which these features are superimposed; it has become customary to identify this main body with the group surrounding NGC 4874, mainly because this galaxy lies on the main peak of the diffuse cluster X-ray emission (see, e.g., White et al. 1993); in particular, this is the point of view of CD95. However, the velocity of NGC 4874 is offset from the average velocity of the cluster, and although not uncommon among brightest cluster members, this is an unexpected result if the galaxy is at rest at the bottom of the potential well of the cluster. Therefore it is also possible that NGC 4874 rather lies at the bottom of the gravitational well of *its own group*, similarly to what is observed for other central dominant galaxies (Bird 1994).

The large number of galaxy velocities now available in the Coma cluster field (613, including our new redshift data-set, Biviano et al. 1995a), and the high resolution of the *ROSAT* PSPC X-ray image allows us to examine the structure and kinematics of the central part of the Coma cluster (within 1 $h^{-1}$ Mpc, where h is the Hubble constant in units of 50 km s$^{-1}$ Mpc$^{-1}$), with the aim of understanding if a *main cluster body* actually exists, independently from the two central subclusters. In this paper we also address the issue of subclustering and ongoing infall onto the cluster in a larger cluster region. Some preliminary results from our work have been presented in Biviano et al. (1994), Gerbal et al. (1994), and Durret et al. (1995).

Shortly before this paper was submitted, we received a preprint from CD95 which also carried out a structural analysis of the Coma cluster, with rather different techniques; while we agree on some of their results, we do not come to the same conclusions on what concerns the structure of the central cluster region.

In § 2 we describe the optical and X-ray data used for our analysis; we analyse the structure of the central region of the Coma cluster, as described by the galaxy and X-ray distributions, in § 3, and the cluster kinematics in § 4; in § 5 we extend our analysis to the external regions; the relevant discussion is provided in § 6; finally we summarize our results and give our conclusions in § 7.

## 2. The data

### 2.1. Optical data

We have taken positions, blue magnitudes, and blue−red colors of galaxies in the Coma region from the catalogue of Godwin, Metcalfe & Peach (1983, hereafter GMP). This catalogue is nominally complete up to an isophotal magnitude $b_{26.5} = 20.0$ (we note these isophotal magnitudes $b$ hereafter), in a 2.63 square degree area centered on $\alpha_{1950} = 12^h 57^m.3$, $\delta_{1950} = +28°14'.4$.

We have assembled a total number of 613 galaxies with redshifts in the Coma region defined by GMP: redshifts for 143 galaxies have been obtained by ourselves through observations with the CFHT in May 1993 (Biviano et al. 1995a); another 167 redshifts have been kindly provided by M. Colless in advance of publication (CD95); finally, another 303 redshifts have been taken from the literature (Bothun et al. 1985; Caldwell et al. 1993; Davies et al. 1987; da Costa et al. 1984; de Vaucouleurs et al. 1976; Gavazzi 1987; Giovanelli & Haynes 1985; Hoffman et al. 1989; Huchra et al. 1983; Huchra et al. 1990; Huchtmeier & Richter 1989; Karachentsev & Kopylov 1990; Kent & Gunn 1982; Leech et al. 1988; Lucey et al. 1991; Palumbo et al. 1983; Sullivan et al. 1981; van Haarlem et al. 1993; Wegner et al. 1990; White et al. 1982; Zabludoff et al. 1993). For the compilation of redshifts from the literature, we have made use of the ZCAT catalogue (Huchra et al. 1995). When several redshift measurements were available for the same galaxy, we have chosen the most accurate one.

In Table 1 is indicated the completeness in velocity as a function of the limiting magnitude in the GMP region and in the inner region (circle of 1500" radius), with respect to all galaxies in the field (cols. 2 and 3, respectively).

For the purpose of our analysis, we have extracted several samples from a combination of the GMP photometric catalogue and our redshift catalogue. In order to reach a high value of completeness in velocities, and still have a sufficiently large number of galaxies in the sample, most of our analyses have been performed on a circular region of radius 1500" (at the distance of the Coma cluster, 1500" correspond to 1 $h^{-1}$ Mpc). We choose a circular region to avoid introducing any preferential direction.

**Table 1.** Completeness in velocity as a function of limiting magnitude. $N_v$ and $N$ represent the number of galaxies with velocities and the total number of galaxies respectively.

| Magnitude range | $N_v/N$ | |
|---|---|---|
| | GMP region | 1500" region |
| $b \leq 17.0$ | 282/346 | 122/123 |
| $b \leq 17.5$ | 325/447 | 148/150 |
| $b \leq 18.0$ | 387/607 | 185/190 |
| $b \leq 18.5$ | 453/820 | 224/248 |
| $b \leq 19.0$ | 508/1128 | 259/313 |
| $b \leq 19.5$ | 554/1609 | 291/403 |
| $b \leq 20.0$ | 605/2510 | 316/544 |
| $17.0 < b \leq 19.0$ | 226/782 | 137/190 |
| $17.0 < b \leq 20.0$ | 323/2164 | 194/421 |

From these samples, we have extracted the cluster members. The membership assignment is based on the galaxy velocity when available (galaxies with velocities in the range $3000 - 10000$ km s$^{-1}$, plus one galaxy with a velocity of 2998 km s$^{-1}$, are assumed to belong to Coma), or on the location of the galaxy in the colour-magnitude band defined by Mazure et al. (1988); for more details on this membership assignement see Biviano et al. (1995b).

We have defined and labeled the samples and subsamples that we have used as follows:

- sample $C_{all}$ contains the galaxies members of the cluster in the GMP region, up to the completeness magnitude $b = 20.0$;
- sample $C$ contains members of the cluster within the circle of 1500" radius centered on the GMP center, up to $b = 20.0$;
- subsample $C_{bright}$ contains $C$ members with $b \leq 17$;
- subsample $C_{faint}$ contains $C$ members with $17 < b \leq 20.0$.
- subsample $C_{faint,19}$ contains $C$ members with $17 < b \leq 19.0$.

Finally, the subscript "$v$" indicates the subsamples of galaxies with available velocities.

The numbers of galaxies with and without velocities in these samples are listed in Table 3 (see next section).

### 2.2. X-ray data

X-ray images of Coma obtained with the *ROSAT* PSPC were kindly made available to us by S. White before their release in the Garching public archive. We combined two of these images, taken in the "hard" energy band (0.4-2.4 keV), totalling an exposure time of about 40000 s. The image sizes were 256×256 pixels$^2$, with a pixel size of 15" (i.e. 10 $h^{-1}$ kpc at the distance of Coma).

## 3. The central region of Coma: structure

### 3.1. Methods of analysis

In order to visualize the distribution of galaxies on all scales at once, we have used the method of adaptive kernels (e.g. Silverman 1986). The adaptive kernel technique is widely used in the astrophysical literature (e.g. Pisani 1993, 1995; Merritt & Gebhardt 1994). Here we have also used this technique to draw maps of various moments of the galaxy velocity distribution (see Figs. 2 and 12).

The adaptive kernel technique requires an initial kernel size; this size is then modified according to the local density of points. Unless otherwise stated, we used an initial kernel size equal to 200" (i.e. 0.13 $h^{-1}$ Mpc at the distance of the cluster), since this is close to the intrinsic scale of many structures in the Coma cluster (Mellier et al. 1988; Escalera et al. 1992), and also happens to be nearly equal to the "optimal" kernel size (Silverman 1986) for the sample $C$, which is the most used sample in our analysis.

The significance of the detected structures is estimated through a bootstrap resampling analysis. See Appendix A for further details.

For the X-ray data, we have used the wavelet analysis, which easily removes noise and allows by its multi-scale approach to detect (at a $> 3\sigma$ level) both the point-sources and the extended sources. A thorough description of the method we have used is given in Slezak et al. (1994; see also Grebenev et al. 1995), and in Appendix B. The wavelet method has already been applied to the investigation of both the galaxy and gas distributions in the Coma cluster (Escalera et al. 1992; Gerbal et al. 1994; Vikhlinin et al. 1994).

### 3.2. The galaxy distribution

We display in Fig. 1 (upper panel) an adaptive kernel density map of all the cluster members in the circle of 1500" radius (i.e. sample $C$); this and the following adaptive-kernel maps, have 128×128 pixels. The significance level of the features visible in this map has been evaluated through a bootstrap resampling technique; the density-map at $-3\sigma$ confidence level (see Appendix A) is also plotted (lower panel): most features of the original map are still visible in this map.

A complex structure is observed, with a main peak corresponding to the central dominant galaxy NGC 4874, an extension to the East corresponding to the brightest cluster member, NGC 4889, and a less significant extension to the South-East in the direction of another bright galaxy, NGC 4911 (these structures have already been pointed out by Mellier et al. 1988 and Escalera et al. 1992). This map therefore naturally suggests the presence of subclustering.

A classical test for estimating the significance level of the presence of subclustering is that of Dressler & Shectman (1988). In this test one computes the local mean ve-locities and velocity dispersions of all possible groups of 11 neighboring galaxies, and the deviations, $\delta$'s, of these local values from the global cluster mean velocity and velocity dispersion (see also Appendix A). Large $\delta$ values indicate a high probability for the cluster to contain substructures. Bird (1994) suggests that using $\sqrt{N_v}$ instead of 11 galaxies in the evaluation of the $\delta$ parameters (where $N_v$ is the total number of galaxies with available velocities), maximizes the sensitivity of the test to significant structures while reducing sensitivity to noise fluctuations.

We have run this test (modified according to the suggestion of Bird 1994) on our sample $C_v$, which contains 255 galaxies with available velocities, and found a probability of 0.921 that substructures are present (based on 1000 Montecarlo simulations). In order to visualize these substructures, we have plotted an adaptive kernel $\delta$-map; notice that this is quite comparable to the weighted wavelet method developed by Escalera et al. (1994). We caution the reader that it is not our intention to improve the statistical test for detection of substructure developed by Dressler & Schectman (1988); the only purpose of our technique is to allow a 2D visualisation of the values of the parameter $\delta$ used to identify susbtructures. This is done following the procedure described in Appendix A. Iso-$\delta$ contours for the sample $C_v$ are shown in Fig. 2.

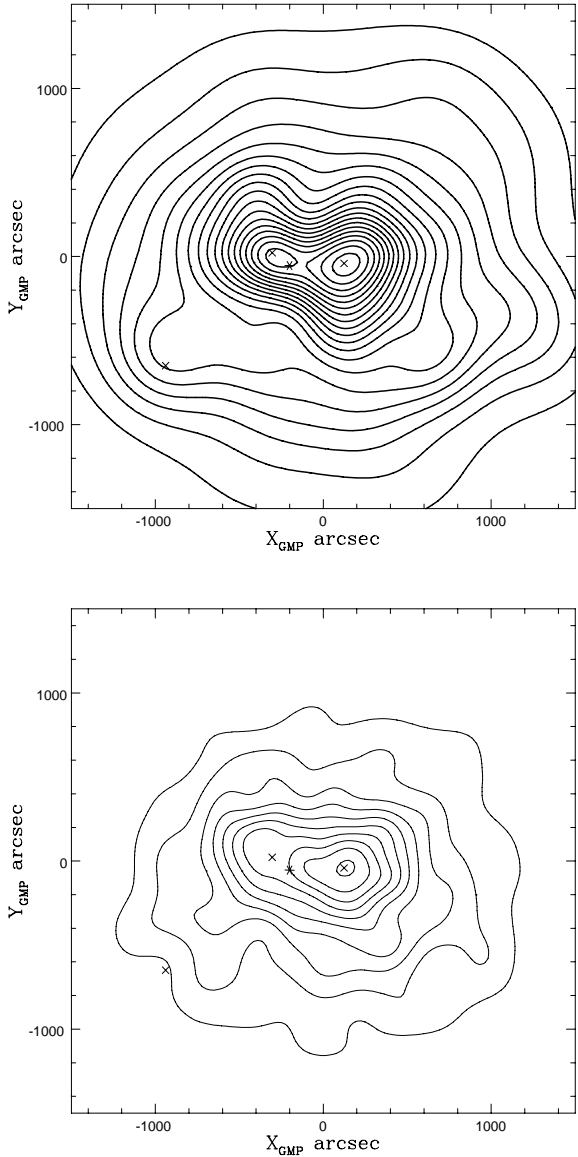

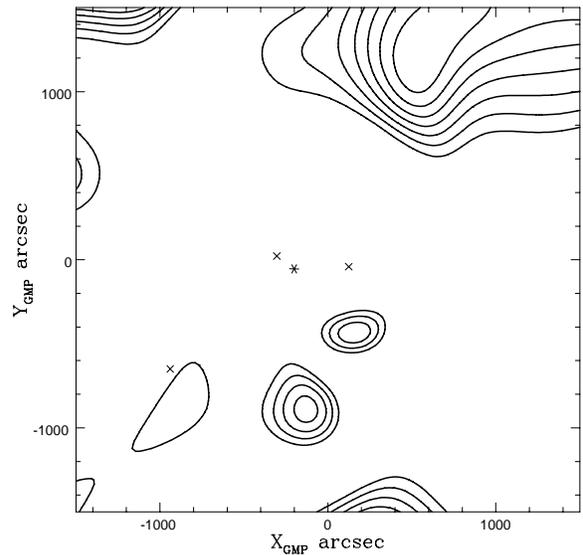

**Fig. 1.** Upper panel: adaptive kernel density map of all the galaxies belonging to the Coma cluster within a circle of radius 1500" (sample $C$). In this figure and in the following ones, coordinates are in arcseconds with respect to the cluster centre defined in GMP; North is to the top and East to the left; isodensity contours are spaced by $1 \times 10^{-5}$ galaxies/arcsec$^2$. The galaxies NGC 4889, NGC 4874 and NGC 4911 at respective positions (-304,22), (124,-41) and (-939,-649) are indicated by x's. The position of a secondary maximum detected in the X-ray map is indicated by an asterisk at position (-201,-54). Lower panel: adaptive kernel density map at $-3\sigma$ confidence level (see Appendix A).

**Fig. 2.** Dressler & Schectman $\delta$-map for sample $C_v$. The average $\delta$ in this map is 40; the first contour starts at a value of $\delta = 40$ and following contours are spaced by an interval of 20 in $\delta$. Note that contours beyond a distance of 1500" from the center are meaningless, because the $C_v$ sample only contains galaxies within this radius. Symbols are the same as in Fig. 1.

One may see on this figure that there are two main peaks in the $\delta$-distribution, corresponding to two regions where galaxy velocities strongly differ from the average velocity field; note that contours beyond 1500" from the center are meaningless, because the sample $C_v$ only contains galaxies within this radius. We have selected the 11 galaxies located within the contours delimiting the two main peaks; these galaxies are listed in Table 2. The 6 galaxies in the first peak have an average velocity of 4518 ± 324 km s$^{-1}$, and the 5 galaxies in the second peak have an average velocity of 8527 ± 265 km s$^{-1}$. Notice that these two values are close to the boundaries of the range of velocities taken for galaxies belonging to Coma. The velocity dispersions of these two groups are more in the range of those of poor clusters than of groups (see e.g., Mazure et al. 1995; Ramella et al. 1995); however, it is possible that the first group is in fact made of 3 galaxy pairs, and that the galaxy with a velocity of 9811 km s$^{-1}$ is not a member of the second group, but an isolated background galaxy. By excluding these 11 galaxies, the test of Dressler & Shectman (1988) gives a probability of 0.529 for the presence of substructures on the remaining 244 galaxies, and the $\delta$-map does not show any other significant feature. So, there are good reasons to believe that these groups are in the fore- and background of the Coma cluster. In the following, we have excluded these 11 galaxies from our samples. The number of galaxies remaining in the subsamples defined in § 2.1 after the elimination of these 11 galaxies, are listed in Table 3.

**Table 2.** Eleven galaxies belonging to the foreground and background groups.

| Group | GMP number | Velocity km s$^{-1}$ | $b$ | (x,y) arcsec |
|---|---|---|---|---|
| 1 | 3262 | 3682± 100 | 16.77 | 68,-417 |
|   | 3275 | 3834± 36 | 18.66 | 76,-419 |
|   | 3383 | 4585± 75 | 18.50 | 178,-394 |
|   | 3400 | 4692± 16 | 15.32 | 191,-310 |
|   | 3473 | 5061± 92 | 18.92 | 249,-409 |
|   | 3522 | 5126± 31 | 16.39 | 300,-266 |
| 2 | 2923 | 8664± 51 | 17.65 | -302,-711 |
|   | 3012 | 8064± 48 | 17.49 | -215,-863 |
|   | 3071 | 8865± 62 | 17.17 | -145,-809 |
|   | 3092 | 8434± 68 | 17.55 | -126,-629 |
|   | 3176 | 9811± 98 | 18.29 | -12,-808 |

In Fig. 3 we have displayed the adaptive kernel density map and the density map at $-3\sigma$ confidence level, for the sample $C$, after the elimination of these 11 galaxies; the map appears slightly smoother than before (compare with Fig. 1), but the most relevant features are unchanged. In particular, the two peaks around the two central dominant galaxies are still visible. It is not surprising that these peaks were not detected by the Dressler & Shectman (1988) test, since the velocity distribution of galaxies within these groups is not very different from the overall cluster velocity distribution.

**Table 3.** Number of cluster members in the various subsamples, after the elimination of 11 galaxies in two fore- and background groups. $N_v$ and $N$ represent the number of galaxies with velocities and the total number of galaxies respectively.

| Sample | $N_v/N$ |
|---|---|
| $C_{all}$ | 469/1372 |
| $C$ | 244/372 |
| $C_{bright}$ | 118/119 |
| $C_{faint}$ | 126/253 |
| $C_{faint,19}$ | 108/141 |

The structure of Coma as visible from Fig. 3 (upper panel) remains complex, despite the elimination of the groups of galaxies in the foreground and background, and raises the natural question of whether there really is a *Coma cluster*, i.e. an underlying main structure. As we discuss in § 6.1, there are reasons to believe that faint galaxies may be more faithful tracers of the main cluster body potential. For this reason, we have split our sample $C$ into the two sub-samples $C_{bright}$ and $C_{faint}$ previously described. We have chosen the magnitude $b = 17$ to separate these two samples because this magnitude corresponds to the dip in the Coma cluster luminosity function (Biviano et al. 1995b), and thus naturally separates bright from faint galaxies in Coma. Moreover, the number of galaxies with available redshifts in the two subsamples is approximately the same.

It turns out that the structures seen in the distribution of bright galaxies are quite different from those seen in the distribution of faint galaxies, as can be seen by plotting the adaptive kernel maps of the $C_{bright}$ (Fig. 4) and $C_{faint}$ (Fig. 5) samples respectively.

As seen in these two figures, the relative importance of the group surrounding NGC 4874 is strongly emphasized in the bright galaxy distribution, while the two peaks around the two giant central galaxies are no longer present in the faint galaxy distribution. The appearance of the map shown in Fig. 5 is more regular, with a single density maximum located between the position of the two central galaxies. This secondary peak is located at GMP coordinates (-201,-59), and therefore at distances from NGC 4874 and NGC 4889 of 323" and 133" respectively. While there still is an overdensity around NGC 4874, it is less prominent than before; this overdensity continuously decreases (and the secondary peak overdensity increases) when the fraction of faint galaxies is increased (i.e. going from sample $C_{bright}$ to $C_{faint,19}$ and to $C_{faint}$).

By looking at the density map at $-3\sigma$ confidence level (lower panel of Fig. 5) one can see that the maximum in the faint galaxy distribution has not a very well determined position and tends to shift towards NGC 4874. We have estimated the uncertainty on the position of this peak by running 1000 bootstrap resamplings on the $C_{faint}$ sample, and estimating the position of the highest peak in each bootstrap sample; the dispersion in the positions of the highest peaks is 315". This large dispersion is probably due to the presence of two density peaks, the one visible in Fig. 5 (upper panel) and the very strong peak around NGC 4874 outstanding in the bright galaxy distribution and not totally erased in the faint galaxy distribution. The striking positional coincidence between the density peak of the faint galaxy distribution, and a secondary maximum detected in the X-rays (see § 3.3), not associated with any of the two central dominant galaxies, strongly argues in favour of the reality of this peak.

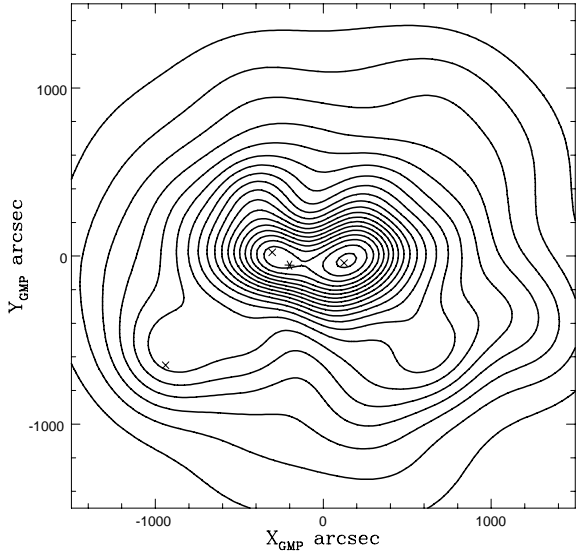

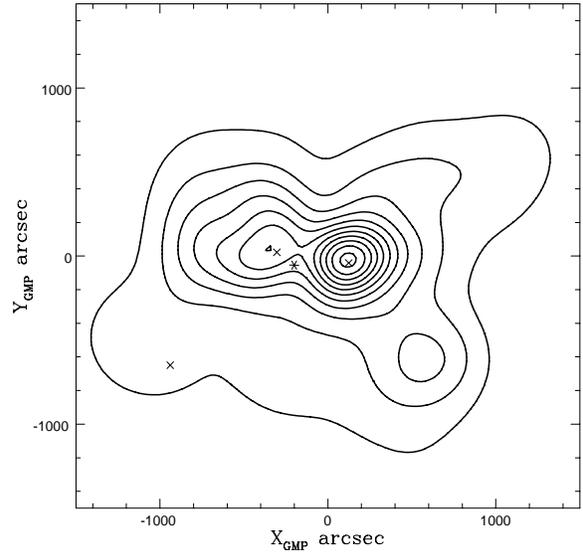

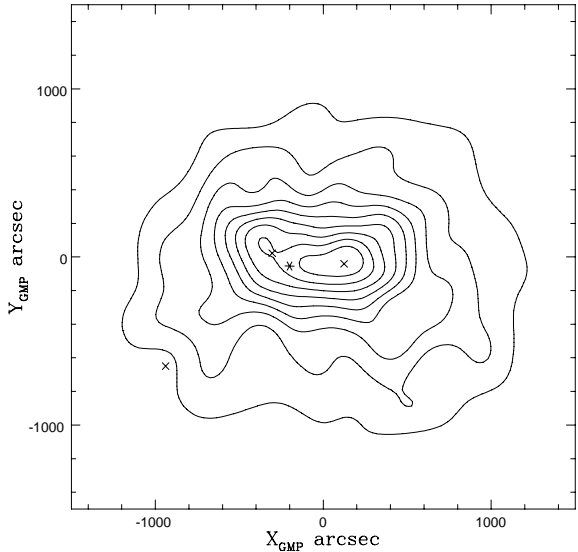

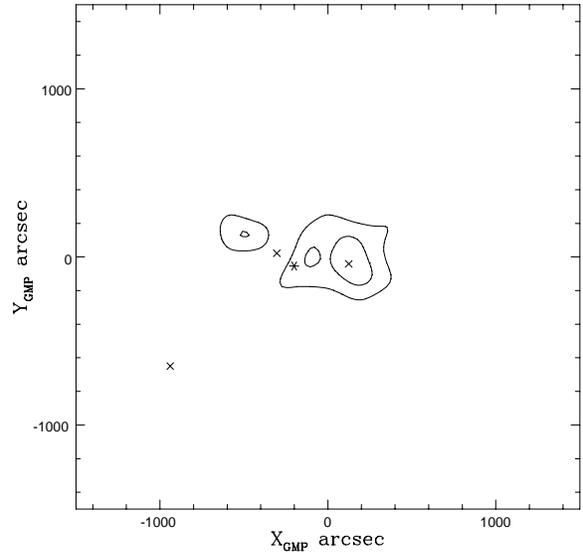

**Fig. 3.** Upper panel: adaptive kernel map of all the galaxies belonging to the Coma cluster within a circle of radius 1500" (sample $C$), after the elimination of the 11 galaxies belonging to the background and foreground groups. The contour spacing and the symbols are the same as in Fig. 1. Lower panel: adaptive kernel density map at $-3\sigma$ confidence level (see Appendix A).

**Fig. 4.** Upper panel: adaptive kernel map of the galaxies belonging to the Coma cluster with magnitudes $b \leq 17.0$ (sample $C_{bright}$). The contour spacing and the symbols are the same as in Fig. 1. Lower panel: adaptive kernel density map at $-3\sigma$ confidence level (see Appendix A).

Magnitudes therefore appear as an efficient tool of discrimination: the cluster appearance changes when selecting samples of luminous or faint galaxies. Bright cluster galaxies are mainly located around the brightest members, while faint ones delineate a smoother single peaked structure.

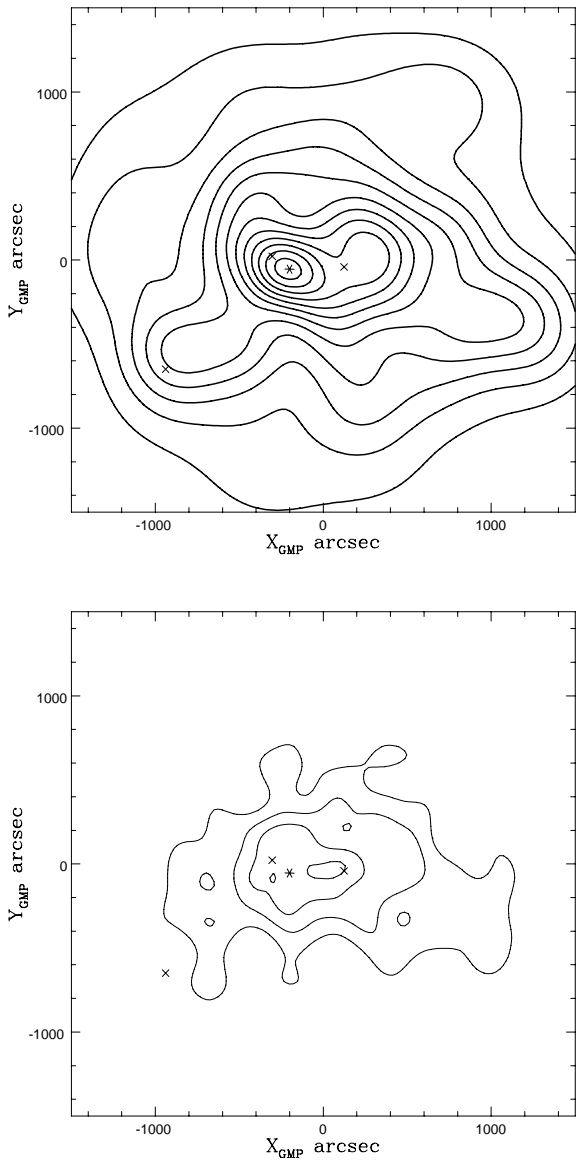

**Fig. 5.** Upper panel: adaptive kernel map of the galaxies belonging to the Coma cluster with magnitudes $b > 17.0$ (sample $C_{faint}$). The contour spacing and the symbols are the same as in Fig. 1. Lower panel: adaptive kernel density map at $-3\sigma$ confidence level (see Appendix A).

### 3.3. The X-ray distribution

We have performed a wavelet analysis on the whole X-ray image of Coma, but for the purpose of comparison with the optical sample $C$, we concentrate here only on the central part of this image (within a radius of $\sim 1500''$). The reconstructed X-ray image (i.e. that obtained after subtracting the noise and artefacts, following a technique described in Appendix B) is displayed in Fig. 6. The four wavelet images, obtained from the noise-subtracted image through the wavelet transform, corresponding to wavelet scales of 15", 30", 60" and 120" (i.e. 10, 20, 40 and 80 $h^{-1}$ kpc at the distance of the cluster, respectively) are displayed in Figs. 7 to 10.

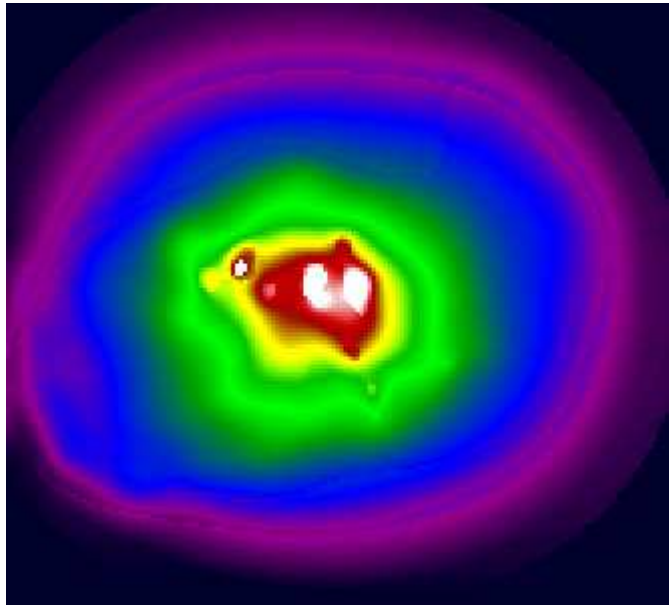

**Fig. 6.** X-ray image obtained after removing the noise and the artefacts (see Appendix B).

In the two first images (Figs. 7-8), we find significant ($> 3\sigma$) unresolved (i.e. not larger than one pixel) features. These generally have optical counterparts which would correspond to X-ray emitting galaxies. We find 12 galaxies in our field (see Table 4). Our analysis is model independent, and allows the detection of X-ray sources. The method used by Dow & White (1995, a fitting procedure) gives actual measurements of the X-ray luminosities, but they only detect 4 galaxies in X-rays, and give upper limits for the others. Note that NGC 4874 and NGC 4889 are rather strong X-ray emitters (see Table 4). The strong X-ray source visible on Figs. 7-8 (labelled with a "Q") is a quasar.

Structures are also observed around the positions of the two central galaxies at larger wavelet scales (see Figs. 9- 10). These detections clearly indicate the existence of two X-ray emitting groups around these two galaxies, in addition to the X-ray emission due to the galaxies themselves. Vikhlinin et al. (1994) have already found these groups in the same *ROSAT* image, by applying a wavelet transform but at a single medium scale.

The present analysis reveals two other features of particular interest:

1. A third maximum (labelled "A") which is clearly visible in Figs. 8-9 between the two maxima associated with NGC 4874 and NGC 4889; this feature is not re-lated to any particular galaxy and was also noted by White et al. (1993). It is located at GMP coordinates (-201,-54), which almost coincides with the position of the density maximum in the faint galaxy distribution (the distance between the two peaks is only 5"). This X-ray maximum is indicated by an asterisk in the adaptive-kernel maps.

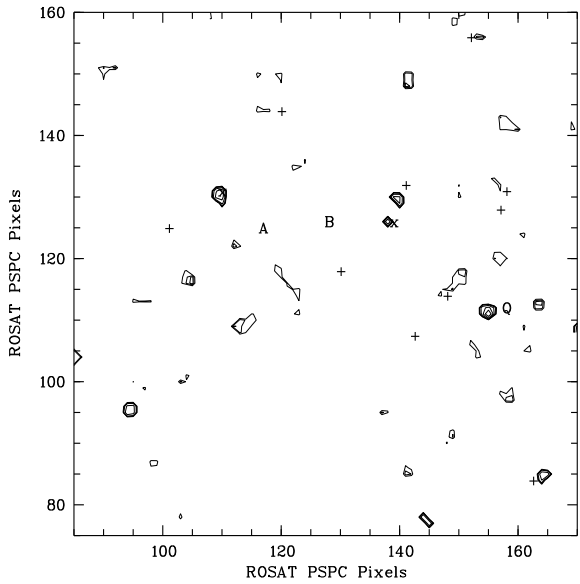

**Fig. 7.** Wavelet image at a scale of 1 pixel (15"). The isophotes are drawn from 0.5 to 7 by steps of 0.5. The X and Y scales are drawn in units of *ROSAT* PSPC pixels of 15". The two brightest galaxies, NGC 4874 and NGC 4889, are indicated by x's. Other labels are: A for the secondary maximum, B for the filament between the two brightest galaxies, and Q for a quasar in the field. The positions of 10 bright galaxies are marked by crosses (see Table 2).

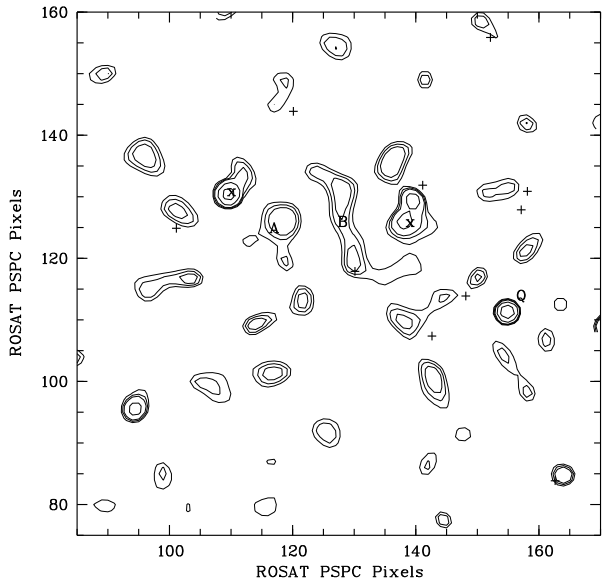

**Fig. 8.** Same as Fig. 7 at a scale of 2 pixels (30").

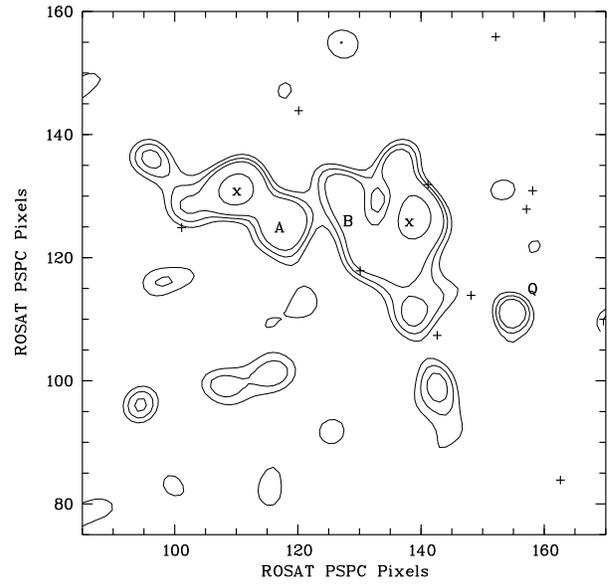

**Fig. 9.** Same as Fig. 7 at a scale of 4 pixels (60").

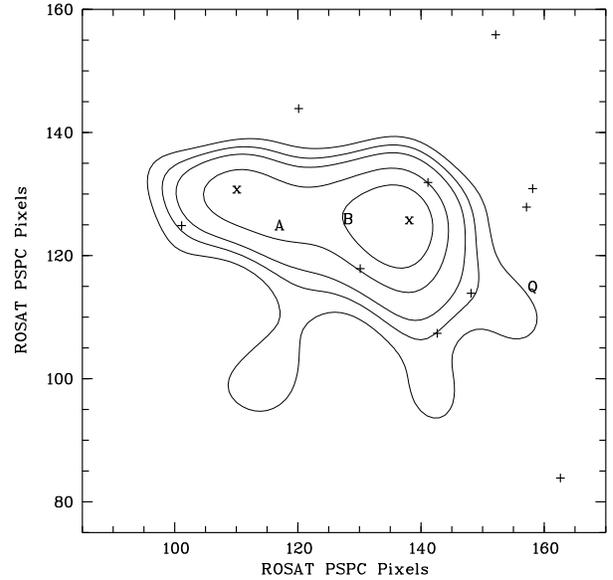

**Fig. 10.** Same as Fig. 7 at a scale of 8 pixels (120").

2. A filamentary structure (labelled "B") between "A" and the galaxy NGC 4874 is visible in Figs. 8-9.

**Table 4.** Optical counterparts to X-ray features detected with the wavelet technique. Col. 1 : galaxy name; col. 2 : magnitude in the GMP catalogue; col. 3 : log $L_X$ in the hard (0.4-2.4 keV) energy band from Dow & White (1995).

| Galaxy name | $b$ | log $L_X$ |
| --- | --- | --- |
| NGC 4874 | 12.78 | 42.19 |
| NGC 4889 | 12.62 | 42.11 |
| NGC 4898 A | 14.85 | 40.91 |
| NGC 4867 | 15.44 | <40.91 |
| NGC 4883 | 15.43 | <40.86 |
| IC 3959 | 15.27 | 40.79 |
| IC 3973 | 15.32 | <40.79 |
| NGC 4869 | 14.97 | <40.76 |
| NGC 4873 | 15.15 | <40.69 |
| NGC 4864 | 14.70 | <40.68 |
| NGC 4865 | 14.54 | <40.62 |
| 1257.3+2758 | 15.15 | <40.47 |

These two features can actually be seen directly on the reconstructed picture (Fig. 6).

### 3.4. Comparison of optical and X-ray maps

The distributions of the galaxies and X–ray gas are remarkably similar. In particular, the two brightest galaxies are surrounded by galaxy groups that are also seen in the X-ray maps, and the faint galaxy population has a single density peak which is also found as a secondary maximum in the X-ray maps (feature "A"). The gas and galaxies seem therefore to be "swimming" in the same gravitational potential. However, there may exist features present on X-ray maps which are absent on optical maps. The response to the gravitational field of these two kinds of matter, which have very different intrinsic physical properties, is different. Notice also that hydrodynamical simulations predict emitting processes from the plasma not directly related to the potential well (Evrard 1990; Roettiger et al. 1993). As we discuss below (see § 6.1), feature "B" may be an example of such processes.

## 4. The central region of Coma: kinematics

As we have seen in § 3.2, the faint galaxies trace a much more regular single-peaked structure than the bright galaxies, which are mostly grouped around the two brightest galaxies. We now make use of the velocity data for investigating the kinematics of these two populations. For this purpose we use the $C_{bright,v}$ and the $C_{faint,v}$ samples. As we have velocities only for 49 % of the galaxies in the $C_{faint,v}$ sample, we have checked the results concerning the kinematics of the faint galaxy population on the sample $C_{faint,19,v}$, which has 76 % of its galaxies with known redshifts.

### 4.1. The velocity gradient

The sample $C_{bright,v}$ is dominated by the two galaxy groups around the central dominant galaxies; this can be seen by splitting the sample in two subsamples, one with velocities $v \leq 6500$ km s$^{-1}$(and therefore closer to the velocity of NGC 4889), and the other with $v > 6500$ km s$^{-1}$. There are 38 and 80 galaxies in the two subsamples respectively. The adaptive kernel density maps of these two subsamples (which are 99 % magnitude complete) are shown in Fig. 11 (upper and lower panel, respectively, for the low and high velocity sample).

On the other hand, the sample $C_{faint,v}$ does not show evidence for the presence of the two groups even if we consider separately the two subsamples at velocities respectively lower and higher than 6500 km s$^{-1}$. Instead, there is evidence for a continuous velocity gradient in the opposite direction to the gradient between the two central galaxies; this is shown in the adaptive kernel map of the average velocity of this sample (Fig. 12; note that we have used an initial kernel size three times as large as that used in the other maps, i.e. 600" in this case, in order to increase the signal-to-noise in this map; see Appendix A for a more thorough explanation). Such a gradient has also been noticed by CD95 (see also Durret et al. 1995). Since this figure is drawn from a sample defined in a circular region, no preferential direction has been artificially introduced.

Since the velocity field of the $C_{faint,v}$ sample of galaxies is complex, we have looked for the direction along which the strongest correlation is obtained between the velocities and the coordinates along this direction, in order to have a quantitative estimate of the position angle and statistical significance of this gradient. The position angle of the maximal correlation is 40° (defined anticlockwise from North), with a rather large uncertainty of ±28°, estimated by performing 1000 bootstrap resamplings of the data. This large uncertainty could be due to the existence of two superposed gradients, one in the East-West and the other in the North-South directions, as suggested by Fig. 12. The velocity gradient along the axis at position angle 40° is 0.25 ± 0.13 km s$^{-1}$ arcsec$^{-1}$ (375±195 km s$^{-1}$ h Mpc$^{-1}$), as deduced from a regression analysis between velocities and coordinates along this axis.

The statistical significance of this gradient computed by the non-parametric Kendall correlation coefficient between velocities and positions is 0.988. Its reality is further confirmed by the analysis of the more complete sample $C_{faint,19,v}$: the direction of the maximal correlation is almost the same, 35° ± 23°, and the statistical significance is even higher, 0.998, despite the fact that the number of galaxies is lower in this sample; the gradient along the axis at position angle 35° is 0.47 ± 0.15 km s$^{-1}$ arcsec$^{-1}$ (705±225 km s$^{-1}$ h Mpc$^{-1}$). This gradient is much larger than the one found before, suggesting that incompleteness in the sample $C_{faint,v}$ has led to an underestimation of the true gradient intensity.

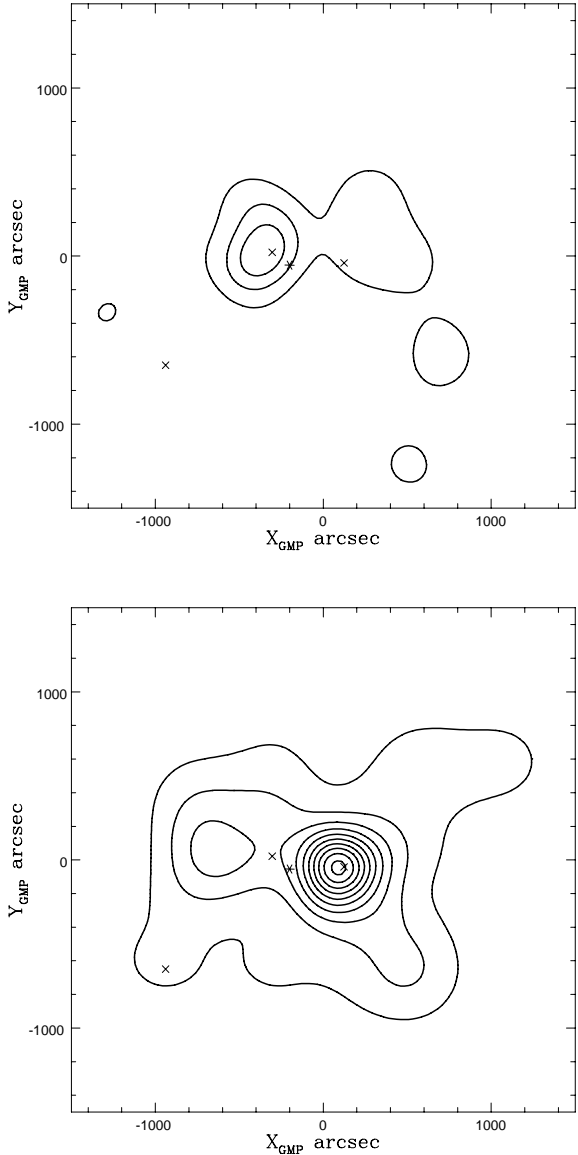

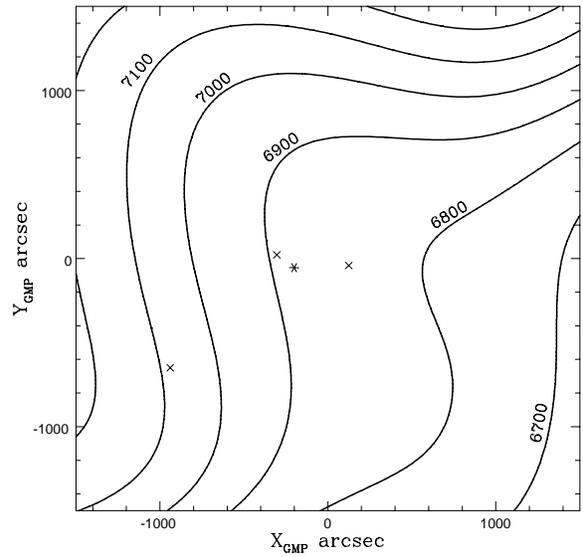

**Fig. 12.** Mean velocity adaptive kernel map for faint galaxies ($C_{faint,v}$ sample). Contour levels are spaced by 100 km s$^{-1}$, and labelled with their mean velocity value. An initial kernel size three times as large as that used in the previous maps has been used here, in order to increase the signal-to-noise in this figure. Symbols are the same as in Fig. 1.

**Fig. 11.** Adaptive kernel density map of the galaxies in the sample $C_{bright,v}$ with $v \leq 6500$ km s$^{-1}$ (upper panel) and with $v > 6500$ km s$^{-1}$ (lower panel). The contour spacing and the symbols are the same as in Fig. 1.

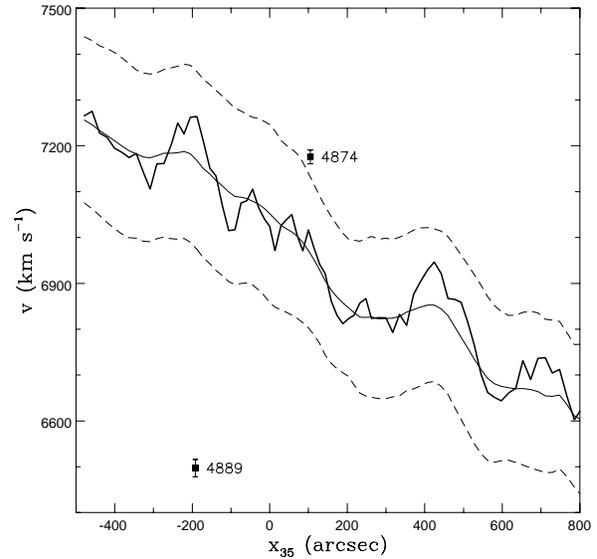

Consistently with the lack of any coherent features in the average velocity maps of the sample $C_v$ and $C_{bright,v}$, we find no significant gradients on these samples, nor do we detect any significant gradient outside this inner circle of radius 1500", consistent with previous findings (Gregory 1975; Gregory & Tifft 1976).

Fig. 13 shows the running velocity profile for galaxies in the sample $C_{faint,v}$ vs. their positions along the direction of the velocity gradient (position angle 35°); the $\pm 1\sigma$ confidence bands are estimated via 1000 bootstrap resamplings. The running velocity profile along the axis at position angle of 40° as well as the running velocity profile of the $C_{faint,19,v}$ sample are similar and are not displayed here. It must be noticed that neither NGC 4874 nor NGC 4889 are located on the running velocity curve, and their relative velocity gradient has an opposite direction to the gradient of the main cluster body.

**Fig. 13.** Running velocity profile (heavy line) of the $C_{faint,v}$ sample along the gradient direction (35°); the dashed lines indicate the $1\sigma$ upper and lower confidence bands as obtained via 1000 boostrap resamplings; the thin line in the middle is the average of the boostrap resamplings. The two brightest galaxies of Coma are indicated by squares; error-bars denote the observational uncertainties on their velocities.

**Table 5.** Characteristics of the velocity distributions

| Sample | mean km s$^{-1}$ | dispersion km s$^{-1}$ | skewness | kurtosis | Tail Index | Omnibus-test probability |
|---|---|---|---|---|---|---|
| $C_v$ | $6901 \pm 86$ | $1140 \pm 56$ | $-0.2 \pm 0.2$ | $0.4 \pm 0.3$ | 1.01 | 0.94 |
| $C_{bright,v}$ | $6827 \pm 105$ | $1119 \pm 81$ | $0.0 \pm 0.2$ | $0.2 \pm 0.4$ | 0.93 | 0.59 |
| $C_{faint,v}$ | $6967 \pm 102$ | $1158 \pm 81$ | $-0.4 \pm 0.2$ | $0.6 \pm 0.4$ | 0.98 | 0.98 |
| $C_{all,v}$ | $6955 \pm 56$ | $972 \pm 35$ | $-0.3 \pm 0.1$ | $0.4 \pm 0.2$ | 1.00 | $> 0.99$ |

### 4.2. Velocity distributions

The histograms and characteristics of the velocity samples $C_v$, $C_{bright,v}$ and $C_{faint,v}$ are displayed in Fig. 14 and Table 5. The values indicated in Table 5 are calculated with a biweight analysis (see, e.g., Beers et al. 1990), and keeping into account the errors on velocities and the reduction to the cluster reference frame (Danese et al. 1980). All mean velocities and values of $\sigma$ are compatible with one another. The values we find for the mean cluster velocity and velocity dispersion are not different from those calculated by CD95.

The velocity of NGC 4874 is $7176 \pm 15$ km s$^{-1}$, a value larger than the average cluster velocity by 275 km s$^{-1}$. The velocity of NGC 4889 is $6497 \pm 19$ km s$^{-1}$, smaller than the mean velocity of $C_v$ by 404 km s$^{-1}$. These results are significant at levels larger than $4\sigma$. The two central dominant galaxies are essentially in a symmetrical situation (within $\approx 100$ km s$^{-1}$) relatively to the $C_v$ mean velocity.

In Table 5 we also list the values for the skewness, the kurtosis, and the Tail Index, relative to the Gaussian distribution, and the probability of rejection of gaussianity according to an omnibus test, for the various samples. The Tail Index (see, e.g., Beers et al. 1991) is a robust measure of the tails in a distribution, being lower than unity when the tails are underpopulated relative to a Gaussian. The omnibus test makes use of both the skewness and the kurtosis to provide the probability that a given distribution is significantly not gaussian (see, e.g., D'Agostino 1986).

Using a range of statistical tests, CD95 have shown that the velocity distribution *of the whole GMP field*, departs significantly from a gaussian. This is confirmed by our analysis (see Table 5, sample $C_{all,v}$). On the other hand, the galaxy velocity distribution in the central cluster region (sample $C_v$) is only slightly non-gaussian. This non-gaussianity mainly arises from the faint galaxy population. In fact, the $C_{bright,v}$ velocity distribution is not significantly different from a Gaussian, even if the low value of the Tail Index suggests the superposition of two popula-tions with different mean velocity in the data-set, probably the two groups around the central dominant galaxies (see Fig. 11 in § 4.1). On the other hand, the $C_{faint,v}$ velocity distribution is significantly non-gaussian, mainly because of its skewness. It is remarkable that if we subtract the average gradient (see § 4.1) from the observed velocities of this sample, the resulting velocity distribution is closer to a gaussian (the omnibus test rejects gaussianity with a probability of only 0.92). This is illustrated in Fig. 14.

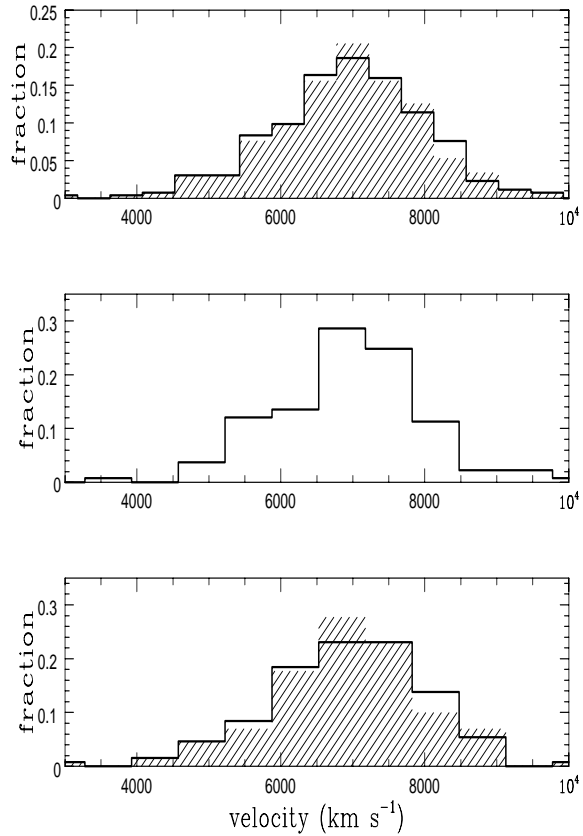

**Fig. 14.** Histogram of velocities of the sample $C_v$ (upper panel), $C_{bright,v}$ (middle panel) and $C_{faint,v}$ (lower panel) before (heavy-line histograms) and after (shaded histograms) the subtraction of the average velocity gradient from the observed velocities of galaxies in the $C_{faint,v}$ sample (which is also part of the $C_v$ sample). The number of bins is approximately equal to the square-root of the number of galaxies in each sample.

Table 6. Properties of 13 BCMs and their "groups"

| GMP number | Other name | x (arcsec) | y (arcsec) | b (mag) | $v_{BCM}$ (km s$^{-1}$) | $N_v/N$ | $<v> - v_{BCM}$ (km s$^{-1}$) |
|---:|---:|---:|---:|---:|---:|---:|---:|
| 2921 | NGC 4889 | -304 | 22 | 12.62 | 6497± 19 | 19/27 | 64±238 |
| 3329 | NGC 4874 | 124 | -41 | 12.78 | 7176± 15 | 30/37 | -208±208 |
| 4928 | NGC 4839 | 1877 | -1694 | 13.51 | 7317± 20 | 9/10 | 67±297 |
| 4822 | NGC 4841 | 1754 | 1828 | 13.88 | 6784± 23 | 1/5 | − |
| 2374 | NGC 4911 | -939 | -649 | 13.91 | 7956± 10 | 7/14 | -552±436 |
| 6523 | NGC 4789 | 4362 | -3220 | 13.94 | 8354± 28 | 1/4 | − |
| 756 | NGC 4944 | -3236 | 783 | 14.00 | 7111± 20 | 0/4 | − |
| 5568 | NGC 4816 | 2828 | -795 | 14.08 | 6948± 24 | 1/3 | − |
| 1176 | NGC 4931 | -2591 | 229 | 14.31 | 5849± 70 | 3/5 | − |
| 2987 | NGC 4892 | -238 | -3860 | 14.70 | 5898± 70 | 0/5 | − |
| 5006 | ZW 160038 | 2010 | 3875 | 14.84 | 7457±100 | 1/5 | − |
| 1900 | ZW 160102 | -1555 | 3866 | 14.91 | 7099± 6 | 1/2 | − |
| 5886 | IC 3900 | 3257 | -2574 | 14.95 | 7115± 25 | 0/4 | − |

## 5. The outer regions of Coma

### 5.1. Other groups

Mellier et al. (1988) claimed the existence of a population of galaxies bound to several of the brightest members of the Coma cluster. We have repeated here their analysis using our larger data-set (see also Biviano et al. 1994). We have selected 13 galaxies with $b \leq 15$ (BCMs hereafter, for Brightest Cluster Members) which are located near a density enhancement, as visible in an adaptive kernel map of the $C_{all}$ sample (Fig. 15). The names of these BCMs, positions, magnitudes and velocities are listed in Table 6. While some of the peaks visible in Fig. 15 may not correspond to real structures, the purpose here is to select all *possible* groups, and to run a statistical test for the presence of two populations within these groups.

We have constructed 13 "groups" centered on these BCMs by taking all the galaxies closer than 250" from each BCM; the radius of 250" has been chosen as a compromise between the need of selecting galaxies close enough to the BCMs, and still having sufficient statistics. Note that with such a radius, the "groups" of NGC 4874 and NGC 4889 would be superposed; we have chosen to assign to the NGC 4874 "group" the galaxies falling in the superposition zone, since from Fig.4, NGC 4874 seems to sit on a larger density peak.

In Table 6 we list in col.(7) the ratio between the number of galaxies with available velocities, $N_v$, and the total number of galaxies, $N$, in each "group", BCMs excluded. Since we do not have any galaxy with available velocity in the vicinity of 3 BCMs, our analysis is effectively restricted to 10 putative "groups". We also give the average velocities for 4 "groups" with enough galaxy velocities available; these average velocities (relative to the velocity of the BCM) are listed in col.(8) of Table 6 (these averages are computed without including BCMs). The average "group" velocities are equal to the BCM velocities, within the errors; however, these errors are quite large. In order to check for the existence of a population of galaxies bound to the BCMs, we have merged together the velocities (relative to the velocities of the BCMs) of the 74 galaxies assigned to the 10 "groups" (BCMs excluded). The null hypothesis that we are testing is that this distribution is the same as that of all the cluster galaxy velocities relative to the cluster mean. This should be the case, if the sample of galaxies in the vicinity of BCMs is a random sample of the cluster population.

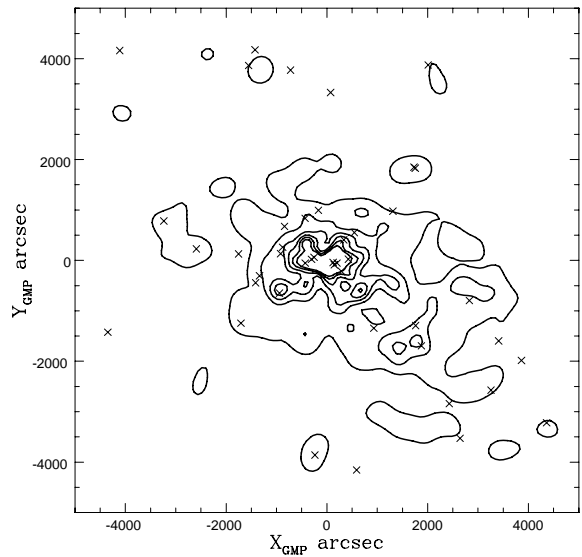

Fig. 15. Adaptive kernel density map of all the cluster members in the GMP region (sample $C_{all}$). The x's indicate all cluster members brighter than $b = 15$ in this field. Contour spacing is the same as in Fig. 1, but contours in the central region have been omitted for clarity.

A Kolmogorov-Smirnov test (see e.g. Stephens 1986) does not reject the hypothesis that the 74 "group" galaxy velocities (relative to their BCM velocities) are drawn from the same parent distribution of the 469 velocities (relative to the cluster average velocity) in the sample $C_{all,v}$. However, there is a difference between the velocity distributions of the "group" galaxies and of the $C_{all,v}$ galaxies, when only galaxies with $b \leq 15.5$ are considered (there are 13 and 89 galaxies brighter than this magnitude, in the "group" and the $C_{all,v}$ sample respectively). This difference is significant at the 94.01 % confidence level, according to the Kolmogorov-Smirnov test. In fact, the velocity dispersion of the 13 "group" galaxies brighter than $b = 15.5$, with respect to the velocities of their BCMs, is only 424 km s$^{-1}$. We conclude that the population bound to the BCMs, discovered by Mellier et al. (1988), is made of bright galaxies only. A similar result was found in § 3.2 for the central region of Coma: bright galaxies tend to lie in subclusters and faint galaxies trace the overall cluster potential.

### 5.2. The region of NGC 4839

Let us now concentrate on the well known group around the giant galaxy NGC 4839. The presence of this group makes the velocity distribution of the $C_{all,v}$ sample significantly skewed and therefore non-gaussian (see Table 5). This was recently pointed out by CD95, who were able to split the group and the cluster components using the KMM algorithm (see, e.g., Bird 1994); they obtained an average velocity of 7339 km s$^{-1}$ and a velocity dispersion of 329 km s$^{-1}$ for the group. The average velocity is in perfect agreement with the velocity of NGC 4839 (7317±20 km s$^{-1}$).

However, CD95 did not address the possible problem of the incompleteness of their sample; in order to check the robustness of their result, we have considered *only* the galaxies of sample $C_{all,v}$ with magnitudes $b \leq 17$; the completeness of this sample is 82 % (see Table 1). By performing a maximum likelihood fit with two gaussians to the velocity distribution of this sample, we recover (within the errors) the average velocity and velocity dispersion found by CD95 for the group population. The nice agreement with the values found by CD95 suggests that their result is robust, despite incompleteness. Using the ratio between the numbers of galaxies belonging to the two gaussians and assuming that all the likely cluster members without available velocities belong to the cluster and not to the group, we can set a lower limit of 0.17 to the ratio between the number of galaxies in the group and in the cluster.

### 5.3. The distribution of star-forming galaxies

As noted by many authors (Sodré et al. 1989; Zabludoff & Franx 1993; CD95; Donas et al. 1995; Gavazzi et al. 1995), the star-forming and the quiescent galaxy populations in Coma have markedly different velocity distributions. We have separated these two populations by their colours, respectively larger or equal to, and smaller than $b - r = 1.7$. In the sample $C_{all}$ there are 278 blue cluster members, of which 64 with measured velocities, and 1073 red cluster members, of which 384 with measured velocities. The density profiles and velocity distributions of the red and blue Coma cluster members are shown in Fig. 16. The density profile of the blue galaxies is flatter than that of the red galaxies in the center, yet the two profiles are similar beyond 1200" (corresponding to 0.8 h$^{-1}$ Mpc) from the center; the velocity distributions of the two populations are different (0.991 probability, according to a Kolmogorov-Smirnov test), both because of a different mean and a different dispersion (the velocity dispersion of the blue galaxies is 30 % larger than that of the red galaxies).

The sample $C_{all,v}$ being rather incomplete, we have performed a similar analysis on the 82 % complete sample of galaxies brighter than $b \leq 17$, and found approximately the same results.

Caldwell et al. (1993) have identified 17 early-type galaxies with an abnormal spectrum in the Coma cluster. Given the peculiarity of their spectrum, most of these galaxies are thought to have suffered a starburst episode $\sim 1$ Gyr ago (Caldwell et al. 1995). A possible explanation for such a starburst is a recent interaction with the cluster environment. These 17 galaxies share a similar velocity distribution as that of the blue galaxies (see Fig. 16). Their average velocity (7526±92 km s$^{-1}$) and velocity dispersion (297±1050 km s$^{-1}$) are also quite similar to those of the NGC 4839 group, although their spatial distribution is not restricted to the region of NGC 4839. Note that the values of the average velocity and dispersion are different from those given in Caldwell et al. (1993) and in CD95, because here we have used the *biweight* estimator, which is more reliable than the classical dispersion in the case of poor data-sets (Beers et al. 1990). While the average velocity is roughly two standard deviations offset from the average velocity of the NGC 4839 group, and the velocity dispersion is affected by an enormous error, this result is intriguing.

## 6. Discussion

### 6.1. The main cluster body

As shown in the previous sections, the bright and faint galaxies have markedly different distributions and kinematics. Bright galaxies tend to be located in groups dominated by their very brightest members, while faint galaxies delineate a much smoother single-peaked structure. This is true not only for the groups around the two core dominant galaxies, but also for other groups even in the external regions. This is reminiscent of the luminosity segregation already observed by Rood (1969). The dichotomy is naturally not perfect, and there probably are faint galaxies located in groups as well, as suggested by the Western extension (towards NGC 4874) of the maximum in the faint galaxy density map (Fig. 5). The X-ray maps show several maxima; in particular, two are centered on NGC 4889 and 4874, and one is coincident with the faint galaxy distribution density peak. The coincidence of the optical and X-ray emitting gas distributions strongly suggests that the detected features are real.

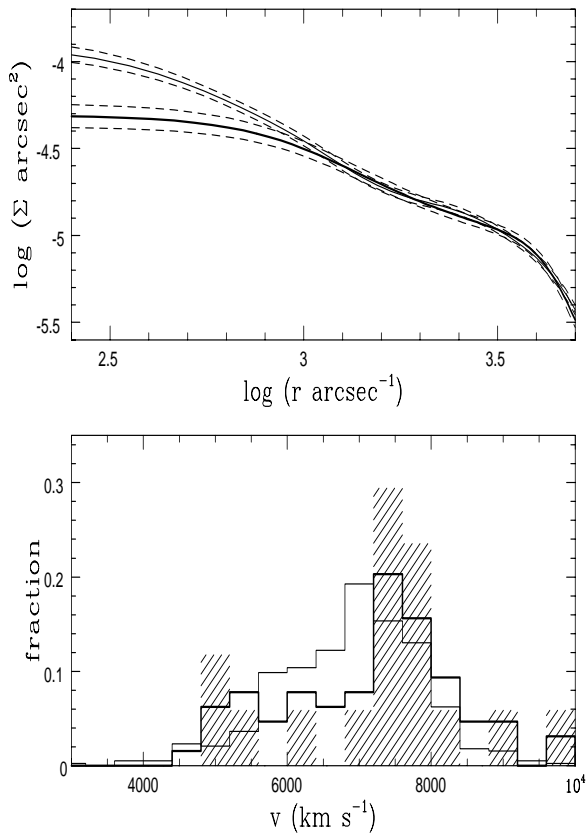

**Fig. 16.** Upper panel: the adaptive kernel number density profiles of blue (thick line) and red (thin line) cluster members. The dashed lines indicate the 1-$\sigma$ confidence bands on the two profiles, obtained with 1000 bootstrap resamplings. Lower panel: the velocity distribution of the blue (thick histogram) and red (thin histogram) galaxy populations, and of the 17 galaxies with abnormal spectra (dashed histogram) detected by Caldwell et al. (1993).

It is usually assumed that the main cluster structure coincides with the overdensity around NGC 4874. However, neither this galaxy nor the brightest cluster member (NGC 4889) are at rest in the cluster potential, since their velocities are different from the average cluster velocity; they are more likely to be located in subclusters still in the process of merging with the cluster. Indeed, the density and X-ray maxima around the two central dominant galaxies indicate the presence of groups linked to these two galaxies; this is supported by the kinematical analysis (see Fig. 11), and by previous investigations (Fitchett & Webster 1987; White et al. 1993; Vikhlinin et al. 1994). CD95 claimed that the two central dominant galaxies do not lie at the mean velocity of their own groups; but they did not realize that the bright galaxies *do* cluster around the two central galaxies, while only the faint galaxies do not (indeed their velocity gradient is opposite to the mean velocity gradient of these two groups, see § 4.1 and § 6.2).

The X-ray secondary maximum (feature "A" in Fig. 9) has already been noted by White et al. (1993), who proposed to identify it with a remnant of gas stripped from NGC 4889. The fact that this peak is also seen in the (faint) galaxy distribution suggests an alternative explanation: *this could be the X-ray maximum of the main body of the Coma cluster.* The "A" feature indeed possesses another property typical of clusters: it is visible on small or medium scales, while it is blurred on the largest scale, indicating that its intrinsic characteristic scale is small, i.e. $\sim 60''$ (corresponding to 40 $h^{-1}$ kpc). Indeed, in two papers devoted to the analysis of 14 X-ray clusters, with improved techniques, Gerbal et al. (1992) and Durret et al. (1994) have shown that X-ray core radii are very small, $\sim 50$–100 $h^{-1}$ kpc (in opposition to what is often claimed in the literature, but in agreement with results derived from the gravitational lensing analysis of the unseen matter distribution). The central peak of the X-ray brightness must then have a medium typical scale, and this is why it is visible on the medium scale image, and not on the larger one.

Our interpretation naturally accounts for the fact that both NGC 4889, and NGC 4874 have velocities significantly different from the cluster mean (see § 4.2), since both galaxies are located within their own groups, eventually falling on to the main body of Coma. These groups would then only be two of the groups merging with, or falling onto the main cluster body. In this case, Coma would still be undergoing a formation process, with the main body continuously accreting matter, i.e. more or less large groups of galaxies. The main body would in fact be formed by the mixing of such groups and clusters arrived in the past, suggesting the continuous existence of violent relaxation processes. Notice that such processes are nothing but *secondary infall*, adapted to a more realistic view of the universe than the simple case of spherical symmetry.

In particular, we suggest that the infall of the NGC 4874 group onto the high density cluster core (region "A" in Fig. 9), is producing a compression wave that has increased the pressure, and, as a consequence, the X-ray luminosity of the gas along the east edge of this group; the X-ray emission of this compressed gas would correspond to the extended filament "B" in Fig. 9. Such an interpretation finds support in hydrodynamical simulations (see e.g. Evrard, 1990) showing that the X-ray emission is not only associated with a smooth gas distribution, but can also arise from "shocks" induced by the arrival of groups into the cluster.

Finally, the fact that we detect the main body in the faint galaxy population, while the bright galaxies are mainly located in subclusters, can be explained as the effect of dynamical evolution in the groups before their infall onto the cluster. The very existence of such large galaxies like NGC 4889 and NGC 4874 argues in favour of this hypothesis. In fact, brightest cluster members are expected to form in groups before violent relaxation occurs (Merritt 1984), probably through mergers which imply small velocity encounters. Such small velocities may be attained among bright, massive galaxies by the loss of their kinetic energy through the process of dynamical friction (see, e.g., Biviano et al. 1992; Yepes & Domínguez-Tenreiro 1992). The groups could then evolve to a core-halo structure, with bright galaxies in their core, and faint galaxies in their halo. When the infall onto the cluster occurs, the less bound halo population would be more easily tidally stripped off the group, and become part of the cluster population, at variance with the core population which could resist longer to tidal disruption. Of course there are probably unevolved groups as well infalling onto the cluster, but these would loose their identity very soon because of tidal disruption, and escape detection; the dense cores of evolved groups would survive longer, as indicated by numerical simulations (González-Casado et al. 1994).

### 6.2. Does the main body rotate?

As shown in § 4.1, the sample $C_{faint,v}$ has kinematical properties on its own; this lends further support to the hypothesis that the density peak of this sample is independent from the density peaks associated with the two central dominant galaxies. In particular, we have shown that the faint galaxy velocity distribution has a significant gradient (see Fig. 12 and 13). This gradient was also noted by CD95, but they did not recognize that it is restricted to the faint galaxy population (no significant gradient is found in the velocity distribution of the bright galaxy population; if anything the gradient is in the opposite sense, see Fig. 11).

It is quite remarkable that this velocity gradient is globally oriented from the North-East to the South-West (35°–40°), in rough coincidence with several main directions in Coma. These privileged directions have recently been pointed out by West (1994), who explains them as a consequence of the infall of structures onto the cluster from the large scale structure filaments surrounding it. In particular, the direction of the velocity gradient is closer to the direction between the centre of Coma and NGC 4839, rather than to the direction to the neighboring cluster Abell 1367. Note, however, that the NGC 4839 group average velocity ($\sim$ 7400 km s$^{-1}$) is opposite to what expected from an extrapolation of the central gradient to the group position.

This gradient is an intrinsic property of the main cluster body. In fact, it cannot simply reflect the presence of subclustering around the two central dominant galaxies, because it goes in the opposite sense. On the other hand, the location of the Coma cluster at the intersection of several sheets may suggest that the velocity gradient reflects the large-scale-structure galaxy distribution. In particular, Coma is within the *Great Wall* (Geller & Huchra 1989; Ramella et al. 1992), which is inclined with respect to the line of sight. However, while in the same direction, the intensity of the gradient found here is much higher than that of the Great Wall, which is indeed roughly similar to the one previously detected by Gregory & Tifft (1976; $\sim$ 0.1 km s$^{-1}$ arcsec$^{-1}$, i.e. $\sim$ 15 km s$^{-1}$ h Mpc$^{-1}$). Morevoer, we only detect a gradient in the inner circle of 1500" radius.

What is the physical meaning of this gradient? The velocity map (see Fig. 12) is complex and not as expected from simple rotation. Moreover, if we interpret the gradient as arising from rotation, the rotational velocity would be too low to account for the flattening of the main body, as one can estimate using the anisotropy parameter $(v/\sigma)^\star$ defined in, e.g., Ferguson & Binggeli (1994, eq.5). The fact that the gradient is principally oriented towards NGC 4839 may suggest a transient tidal effect. Among the numerous consequences of such a transient tidal effect, are the deformation of the matter distribution, and a mean velocity field which is no longer constant. The effects of tides on spheroidal galaxies infalling onto our own Galaxy are described in a paper by Piatek & Pryor (1995). They find that tides deform the spheroidal galaxies and induce apparent rotation: tides produce large ordered motions rather than large random motions. We think that something similar may be happening in the Coma cluster, with the central part of the main body suffering from the tidal influence of the infalling groups, like those around NGC 4874 and NGC 4839. The search for this kind of effect with realistic numerical simulations of clusters could give constraints on the relative masses of the satellite and main body, as well as on the history of encounters.

### 6.3. Ongoing infall

The groups around NGC 4874 and NGC 4889 are likely to be orbiting within the Coma cluster, and they are not at rest at the bottom of the cluster potential. There are at least two more groups, one centered on the galaxy NGC 4839 and the other, less conspicuous, on the galaxy NGC 4911, and other groups may be present around other bright galaxies (Mellier et al. 1988). The average velocities of these galaxies and their groups are also quite different from the cluster mean velocity. Altogether, this suggests that the Coma cluster is still in the process of accreting groups from the surrounding large scale structure.

Not only is Coma accreting groups but also individual galaxies. The flatness of the density profile of the blue galaxy population near the center indicates that these galaxies tend to avoid the center, yet the similarity of their density profile to that of red galaxies (see Fig. 16) at large radii suggests that these blue galaxies *are* cluster members. It is then possible that these galaxies spend only a little time in the central region of the cluster, because they have very elongated orbits. This is also indicated by the double peaked velocity histogram of the blue galaxy population, typical of galaxies on strongly radial orbits (see, e.g., Merritt 1987). This situation is not at all unusual in galaxy clusters (Moss & Dickens 1977; Sodré et al. 1989; Zabludoff & Franx 1993; Biviano et al. 1995c). Supporting this interpretation is the similarity of the blue galaxy velocity distribution to those of the HI-deficient galaxies (see Fig. 3 in Gavazzi 1987), and of the abnormal-spectrum early-type galaxies (see Fig. 16) detected by Caldwell et al. (1993). Both the HI-deficient galaxies and the abnormal-spectrum early-type galaxies are likely to have interacted with the cluster environment, and are therefore probable cluster members, so that this may also be true of the general blue galaxy population.

The group around NGC 4839 has an average velocity close to the peak in the velocity distribution of the abnormal-spectrum galaxies. This fact led CD95 to conclude that both the group and these galaxies are falling onto the cluster coming from the same galaxy filament. While this is certainly a possibility, supported by the similarity of the velocity distributions of the abnormal-spectrum and blue galaxies, the abnormal-spectrum galaxies also have a similar velocity dispersion as the NGC 4839 group, and most of them are located near this group. The possibility then remains that these galaxies are (or have been) members of the NGC 4839 group. In this case, if they have already interacted with the cluster, this would be true of the NGC 4839 group as well.

Burns et al. (1994) suggested that the NGC 4839 group has passed through the core of Coma following a straight trajectory. As shown by González-Casado et al. (1994), this would cause the group structure to be severely affected by tidal forces. However, the NGC 4839 group has a richness, velocity dispersion, and X-ray luminosity that fits very well the relations between these quantities, as derived for clusters of galaxies (Edge & Stewart 1991; Girardi et al. 1993); this argues against the possibility that the group has suffered strong tidal effects. A possible alternative scenario is a non-radial infall of the NGC 4839 group onto the cluster; one or several passages through the outer parts of the cluster may already have occurred, which would leave the group intact, except for some galaxies that may have been lost in its trail. Such a scenario agrees with predictions derived from recent numerical simulations in which the effects of dynamical friction are included (see e.g. Chan et al. 1995 and references therein).

## 7. Conclusions

The analysis of Coma simultaneously in X-rays and at optical wavelengths allows us to draw a coherent picture of this cluster:

1. A main body centered on the peak of the distribution of faint galaxies, which coincides with a secondary X-ray peak, is unveiled; groups are falling on to this main body;
2. Faint galaxies are better tracers of the cluster structure than bright galaxies, since many of the latter are subclustered around the very brightest galaxies;
3. We detect a gradient in the velocities of the faint galaxies tracing the main cluster body; we suggest that this gradient arises from tidal interactions of the main cluster body with infalling groups, in particular with NGC 4839;
4. The infall of the group surrounding NGC 4874 may have induced a pressure wave in the X-ray emitting gas;
5. Groups are still visible mainly because of a bound population of bright galaxies, which has probably evolved to a dense core structure before the infall of their own group, and is thus less prone to tidal stripping;
6. The cluster is still accreting these groups and also individual galaxies from the surrounding large scale structure.

*Acknowledgements.* We deeply thank J.G. Godwin, N. Metcalfe & J.V. Peach for providing us with their unpublished catalogue of stellar objects in the Coma field, M. Colless for providing us with his catalogue of redshifts in advance of publication, and C. Clemens for sending us an electronic copy of the ZCAT catalogue. We are also grateful to S. White for providing us with the *ROSAT* images of Coma before their release in the Garching public archive. We acknowledge the help of D. Fadda in the software implementation of the adaptive kernel technique at an early stage of this project. We thank D. Merritt and A. Pisani for useful discussions. We acknowledge financial support from GDR "Cosmologie", CNRS, and from the European Cosmology Network (a E.U. H.C.M program). C. Lobo is fully supported by the BD/2772/93RM grant attributed by JNICT, Portugal.

## Appendix A: Adaptive kernel maps

A thorough description of the adaptive kernel techniques is given in Pisani (1993, 1995). Here we only address the issues that are not reported in those papers.

### A.1. Statistical significance

In order to assign the statistical significance to a given feature we have run 1000 bootstrap resamplings on each map. In practice, this consists in evaluating an adaptive kernel map of the quantity of interest (typically, the density) for each bootstrap sample; we then compute the standard deviation of the bootstrap values for each pixel in the map. This provides an estimate of the error in the observed quantity in each pixel. We then subtract to the *observed* value in each pixel a given number (say, 3) of standard deviations, and obtain a new map which we compare to the original one, to see if the observed features are still visible. Features affected by large errors will not be visible in the density map at $-3\sigma$ confidence level.

### A.2. Velocity maps

We define an adaptive kernel map of the average velocities by multiplying the 2-dimensional gaussian adaptive kernel, $K_j^{(2D)}(\vec{x})$, of each galaxy $j$, by its velocity (when available), $v_j$:

$$\overline{v}(\vec{x}) = \sum_{j=1}^{N_v} K_j^{(2D)}(\vec{x}) v_j / \sum_{j=1}^{N_v} K_j^{(2D)} \quad (A1)$$

where $\vec{x}$ is the position vector of a given pixel of the image, and $N_v$ is the number of galaxies with available velocity.

In order to increase the signal-to-noise in these maps, we adopt a larger initial kernel size than in the density maps.

### A.3. Substructure maps

Before defining a $\delta$-map, let us remind the definition of the parameter $\delta$:

$$\delta_j = (\overline{v}_j - \overline{v})^2 + (\sigma_{v,j} - \sigma_v)^2 \quad (A2)$$

where $\overline{v}_j$ is the average velocity and $\sigma_{v,j}$ is the velocity dispersion of the galaxy group defined by the $\sqrt{N_v}$ nearest neighbours to the $j$-th galaxy, and $\overline{v}$ and $\sigma_v$ are the average velocity and velocity dispersion of all the $N_v$ galaxies with available velocities.

We then define a local average velocity $\overline{v}(\vec{x})$, as in eq. (A1), and a local velocity dispersion, $\sigma_v(\vec{x})$, as follows:

$$\sigma_v^2(\vec{x}) = \sum_{j=1}^{N_v} K_j^{(2D)}(\vec{x})(v_j - \overline{v}(\vec{x}))^2 / \sum_{j=1}^{N_v} K_j^{(2D)} \quad (A3)$$

Then, as a straightforward extension of the Dressler & Shectman (1988) test, we compute the local values of $\delta$:

$$\delta(\vec{x}) = (\overline{v}(\vec{x}) - \overline{v})^2 + (\sigma_v(\vec{x}) - \sigma_v)^2 \quad (A4)$$

The isopleths of $\delta(\vec{x})$ give the $\delta$-map.

## Appendix B: Wavelet maps

Images which exhibit features of different characteristic sizes at various locations are much more accurately analyzed using space-scale techniques than monoscale informations. In Slezak et al. (1994), we discussed the ability of the wavelet transform to locate these structures according to their typical scales and presented quite extensively the basic mathematical formulae as well as a fast computation algorithm. Therefore, we focus especially this appendix on the use of a multiscale vision model for restoring an image where some components have been subtracted. We have already applied these concepts to design a hierarchical filter which allows to remove the small-scale noise while keeping the relevant details associated to the objects (Slezak et al. 1994). Both improvements and generalization of this method are explained here in a concise way since a full description of the model we use and numerical simulations and tests will be given in Rué & Bijaoui (1995).

### B.1. The multiscale vision model

Using an isotropic analyzing wavelet without any decimation from one scale $a_i = 2^i$ to another, a $2D$ image is transformed into a set of wavelet images $\{W_i\}$ exhibiting details related to different spatial frequencies, so that a complex object showing features of different sizes is associated to non zero coefficients at successive scales. An object now has to be described in this wavelet space and by consequence the connectivity property which defines it in the direct space needs to be enlarged to the $3D$ set of wavelet coefficients $W_i(k,l)$. To do so, one must define an interscale connectivity relation linking the various components corresponding to the connected regions $D_i(n)$ detected at each scale $a_i$ by the field labelling $L_i(k,l) = n$. This image segmentation permits to select significant wavelet coefficients from noisy data. Rué & Bijaoui (1995) decided to base this relationship $\mathcal{R}$ on the location $(k_{max}, l_{max})$ of the maximum value inside $D_i(n)$, and we have adopted the same rule:

$$D_i(n) \; \mathcal{R} \; D_{i+1}(m) \quad \text{if} \quad L_{i+1}(k_{max}, l_{max}) = m \; . \quad (B1)$$

Starting from the largest scale to the smallest one, the identification of the fields which appear to be connected in this manner results in a connection tree. So, any complex object is viewed as an arborescent structure $\mathcal{V}$ in the wavelet space, the whole information of which is required to characterize correctly the object with all its components. A key issue in this approach is thus the possibility to extract easily these components by considering their related sub-trees, i.e. the regions of the wavelet structure connected to the field identified as the largest structural element of the component. One can therefore either study these well-defined components in a separate way or analyze an object where undesirable features like superimposed smaller objects have been quite perfectly removed by eliminating their sub-trees. Note that the partition related to such somewhat bright sub-objects can be distinguished from genuine components belonging to a complex object by the maximum wavelet value inside their root domain. It must be higher than the values of the maxima for the connected fields at the previous larger and next smaller scales since wavelet values indeed increase or decrease according to the match between the wavelet scale and the typical size and amplitude of the signal. The most efficient way to characterize each source detected in the wavelet space is to analyze an image $I$ of this source in the direct space. Let us now address this issue.

### B.2. The image reconstruction

Assuming a Gaussian noise and providing that the standard deviation parameters $\sigma_i$ of the Gaussian density probability distributions for the noise process, $P(W_i(k,l))$, are obtained from the clipped histogram of the oversampled and noise-dominated coefficient map $W_1$ (see Slezak et al. 1994), the fields $D_i(n)$ are labelled by performing an image segmentation at a 3 $\sigma_i$ level on each wavelet map $W_i$. The connection trees are then constructed following the rule described in § B.1. They are pruned in order to remove isolated domains and the existing objects and components are identified from the maximum value of their root fields.

We now have to compute for each object an image $I$ from the set $\mathcal{V}$ of the wavelet coefficients associated to its tree by solving the equation:

$$\mathcal{V} = O(I) = (P_\mathcal{D} \circ WT)(I) \equiv W_\mathcal{D}, \quad (B2)$$

where $WT$ and $P_\mathcal{D}$ design the operators associated to the wavelet transform and to the projection in the domains $\mathcal{D} \equiv \sum_i D_i$ respectively. The mere addition of the thresholded frames is in fact unable to restore an image which will give again the same wavelet coefficients since the thresholding results in a wavelet structure which cannot be considered as the wavelet transform of an image.

This inverse problem has at least one solution: that corresponding to the image which has been processed. Among the class of mathematically acceptable solutions, the choice of the physically correct one is controlled by a regularization condition. Many are available. For instance we first decided to take into account the consistency between $\mathcal{V}$ and $W_\mathcal{D}$, and we used a simple direct algorithm similar to the iterative approach described in Van Cittert (1931) for which the regularization is achieved by constraining the support of the solution $\tilde{I}$ (Slezak et al. 1994). Another choice is to minimize the energy of the residuals between the solution and the initial set of wavelet coefficients, i.e.

$$\|\mathcal{V} - O(\tilde{I})\| \text{ is minimum with } \|\mathcal{V}\| = \Sigma_\mathcal{V} W_i^2(k,l), \quad (B3)$$

which leads to the so-called conjugate gradient algorithm. This method appears better suited for accurate image restoration of partial trees and of sub-trees than our previous technique designed mainly for restoring whole trees, and we decided to apply it to the *ROSAT* X-ray images of the Coma cluster. It is fully described in Rué & Bijaoui (1995), and only the key formulae of their iterative algorithm are given hereafter.

Let us note $\tilde{O}$ the joint operator which transforms a wavelet structure into an image and and $W_\mathcal{D}^{(n)}$ the wavelet transform inside $\mathcal{D}$ of $\tilde{I}$ at step $n$. The joint operator $\tilde{O}$ appears to be basically equal to the addition of the wavelet frames $[WT(W_i)]_i$. We get:

$$\tilde{I}^{(n+1)} = \tilde{I}^{(n)} + \alpha^{(n)} \tilde{I}_r^{(n)} \quad (B4)$$

with $\tilde{I}_r^{(n)} = \tilde{O}(\mathcal{V} - W_\mathcal{D}^{(n)}) + \beta^{(n)} \tilde{I}_r^{(n-1)}$, where $\tilde{I}_r^{(n)}$ stands for the residual image at step $n$ and $\alpha^{(n)}$ and $\beta^{(n)}$ are convergence parameters. At each step, the solution $I^{(n+1)}$ is thresholded in order to get a positive image and the energy of the residual wavelet coefficients $\mathcal{V} - O(\tilde{I}^{(n+1)})$ is tested against a given precision threshold. Initializations are performed by using the classical wavelet reconstruction operator adding all the wavelet planes to the smoothest approximation of the signal for $\tilde{I}^{(0)}$ and by setting $\beta^{(0)}$ to zero.

The reconstruction appears optimal only if the maximum of the wavelet coefficients involved for a given object is included into $\mathcal{V}$. If it is not the case, an important information about the object is missing and the discrepancy between its original and restored images may be large.

Thanks to the unfolding of the data in the wavelet space and by considering whole hierarchical connection trees or part of them, this iterative process enables one to restore either images of objects where the noise and superimposed features have been accurately subtracted without loss of any small-scale information, or images of these features where the extended underlying main objects have been removed.